\documentclass[pra,twocolumn,aps,superscriptaddress,showpacs,floatfix]{revtex4}
\usepackage{amsmath,amssymb,latexsym,amsfonts}
\usepackage{epsfig}
\usepackage{colordvi}
\begin{document}
\title{Fine-scale oscillations in the wavelength- and intensity-dependence of high-harmonic generation: connection with channel closings}
\date{\today}

\author{K. L. Ishikawa}
\email[Electronic address: ]{ishiken@riken.jp}
\affiliation{Integrated Simulation of Living Matter Group, RIKEN Computational Science Research Program, 2-1 Hirosawa, Wako, Saitama 351-0198, Japan}
\affiliation{PRESTO (Precursory Research for Embryonic Science and Technology), Japan Science and Technology Agency, Honcho 4-1-8, Kawaguchi-shi, Saitama 332-0012, Japan}

\author{K. Schiessl}
\affiliation{Institute for Theoretical Physics, Vienna University of 
Technology, Wiedner Hauptstra\ss e 8-10, A--1040 Vienna, Austria, EU}

\author{E. Persson}
\affiliation{Institute for Theoretical Physics, Vienna University of 
Technology, Wiedner Hauptstra\ss e 8-10, A--1040 Vienna, Austria, EU}
\author{J. Burgd\"orfer}
\affiliation{Institute for Theoretical Physics, Vienna University of 
Technology, Wiedner Hauptstra\ss e 8-10, A--1040 Vienna, Austria, EU}

\begin{abstract}
\noindent
We investigate the connection of recently identified fine-scale oscillations in the dependence of the yield of the high-harmonic generation (HHG) on wavelength $\lambda$ of a few-cycle laser pulse [K. Schiessl, K.L. Ishikawa, E. Persson, and J. Burgd\"orfer, \prl {\bf 99}, 253903 (2007)] to the well-known channel closing (CC) effect. 
Using the Lewenstein model of HHG, we identify the origin of the oscillations as quantum interference of many rescattering trajectories. 
By studying the simultaneous variations with intensity and wavelength, different models for the interference of channel closing peaks can be tested. 
Contrary to theoretical predictions for short-range potentials, the peaks are not located neither at nor just below the CC condition, but a significant shift is observed. 
The long Coulomb tail of the atomic potential is identified as the origin of the shift.
\end{abstract}

\pacs{32.80.Rm,42.65.Ky,32.80.Fb} 
\maketitle

\section{Introduction}

\noindent
High harmonic generation (HHG) represents a versatile and highly successful avenue towards an ultrashort coherent
light source  covering a wavelength range from the vacuum ultraviolet to the soft X-ray region \cite{ar:seres05}.
This development has opened new research areas such as attosecond science \cite{Hentschel2001Nature,Tsakiris2003Nature} and nonlinear optics in the XUV region \cite{Sekikawa2004Nature,Nabekawa2005PRL}.
The fundamental wavelength $\lambda$ used in most HHG experiments to date is in the near-visible range ($\sim 800$ nm). 
The cutoff law for the harmonic spectrum $E_c=I_p+3.17U_p$, where $I_p$ denotes the binding energy of the target atom and $U_p=F_0^2/4\omega^2=F_0^2 \lambda^2/16 \pi^2$ the ponderomotive energy ($F_0$: laser electric field strength), suggests that a longer fundamental wavelength would be advantageous to extend the cutoff to higher photon energies, since $U_p$ increases quadratically with $\lambda$. 
This has stimulated an increasing interest in the development of high-power mid-infrared ($\sim 2~\mu$m) laser systems, e.g., based on optical parametric chirped pulse amplification. The first generation of water-window harmonics with clear plateau and cut-off structures has recently been reported \cite{Takahashi2008PRL}.
Along those lines the dependence of the HHG yield on $\lambda$ has become an issue of major interest \cite{ar:tate_scaling07, Schiessl2007PRL, Frolov2008PRL, Schiessl2008JMO, Colosimo2008NP}.
It had long been believed that the spreading of the returning wavepacket would result in a $\lambda^{-3}$ dependence of the HHG efficiency  \cite{ar:lewenstein94} as long as ground state depletion can be neglected \cite{Gordon2005OE}; experimental findings \cite{Shan2001PRA} provided partial support. Recently, however, Tate \emph{et al.} \cite{ar:tate_scaling07} have reported a different wavelength-scaling of HHG between 800 nm and 2 $\mu$m calculated with the time-dependent Schr\"odinger equation (TDSE) for Ar and a strong-field approximation (SFA) for He. They found the yield to be described by a power-law $\propto \lambda^{-x}$  with $5 \le x \le 6$. 
Investigating the $\lambda$ dependence on the level of 
single-atom response for H and Ar by numerically solving the time-dependent Schr\"{o}dinger equation we could confirm the overall scaling with an inverse power law exceeding five \cite{Schiessl2007PRL}, the harmonic yield was found not to depend smoothly on the fundamental wavelength, but to exhibit surprisingly rapid oscillations with a period of $6-20$ nm depending on the wavelength region.
A semiclassical analysis based on the SFA has revealed that the rapid oscillations are due to the interference of five to ten different rescattering trajectories \cite{Schiessl2007PRL}. 
Moreover, we found the oscillations to be stable with respect to variations of the pulse envelope as long as the effective pulse length and thus the number of relevant trajectories remains equal, while the amplitude of the oscillations decreases with decreasing pulse length \cite{Schiessl2008JMO}. 
These observations underscored the view that the oscillations are due to the interference of quantum paths.

Oscillations of the HHG yield have previously been reported in terms of the dependence on the intensity of the driving laser $I_0\propto F_0^2$, both experimentally \cite{Toma1999JPB, Zair2008PRL} and theoretically \cite{Borca2002PRA,Milosevic2002PRA}. Borca {\it et al.} \cite{Borca2002PRA} and Milo\v{s}evi\'c and Becker \cite{Milosevic2002PRA} have shown that HHG is enhanced at channel closings (CC), i.e., when
\begin{equation}
\label{eq:cc}
R=\frac{I_p+U_p}{\hbar\omega},
\end{equation}
is an integer. 
Channel closing in this context refers to the threshold for multiphoton ionization in a laser field. 
Most of these theoretical studies employed zero-range potentials or 
the SFA which both neglect the influence of the long-range potential on the ionized 
electron.

Frolov {\it et al.}\ \cite{Frolov2008PRL} have recently analyzed the wavelength-dependence of HHG in terms of channel closings (or threshold phenomena). 
They have calculated the harmonic yield using the time-dependent effective range theory, and shown that the peaks of the yield oscillation around $\lambda=1$ $\mu$m coincide with integer values of $R$ if an effective ionization potential $\tilde{I}_p$ (e.g., 10.5 eV for H) is used in place of $I_p$ in Eq.\ (\ref{eq:cc}). 
This method is, however, strictly applicable only for short-range potentials and also neglects the excited atomic states. 
On the other hand, we have recently found \cite{Schiessl2008JMO} channel closing peaks in the TDSE-calculated HHG yields around 1 $\mu$m and 2 $\mu$m wavelengths which are characterized by a spacing of $\delta R=1$, as expected from the CC picture, when the true ionization potential $I_p$ is used.

In the present paper, we study the connection between the oscillation in the wavelength-dependence of the HHG yield and the channel-closing in more detail. 
We study the harmonic spectrum in the two-dimensional parameter space of intensity $I_0$ and driver wavelength $\lambda_c$. 
We compare the results of the full 3D TDSE solution with the strong-field approximation and a truncated Coulomb-potential model in order to delineate underlying mechanisms. 
We find that the correspondence of the modulation period to $\delta R=1$ holds for a wide wavelength range between 800 nm and $2 \mu$m, and that the peak positions in terms of $R$ are almost independent of laser intensity. 
The systematic displacement of the peak positions relative to integer values is found to be consistent with the effect of the long-range Coulomb tail on the returning electron. 
 
The present paper is organized as follows.
Section \ref{sec:model} summarizes the two complementary integration schemes employed for a full numerical solution of the TDSE. 
In Sec.\ \ref{sec:wavelength} we discuss the overall wavelength dependence at a fixed value of fundamental intensity all the way from $\lambda=$800 nm to $2\mu$m. 
We also analyze small-scale oscillations in terms of quantum-path interference based on the saddle-point analysis (SPA) \cite{ar:lewenstein94,Milosevic2002PRA}.
In Sec.\ \ref{sec:intensity_dependence} we investigate the variation of the $\lambda$-dependence of HHG with intensity and pulse shape. 
In Sec.\ \ref{sec:modulation_period} we discuss the period of the oscillations in terms of the channel-closing number $R$ and investigate its robustness against the variation of the wavelength region, the driver intensity, and pulse shape. 
In Sec.\ \ref{sec:truncated_Coulomb} we discuss the origin of the peak shift from integer $R$ values and clarify how the Coulomb tail of the atomic potential affects the peak position. Conclusions are given in Sec.\ \ref{sec:conclusions}. 
Atomic units are used throughout the paper unless otherwise stated.

\section{Numerical Methods}
\label{sec:model}

We solve the atomic time-dependent Schr\"{o}dinger equation (TDSE) in the length gauge for a linearly polarized laser field with the central wavelength $\lambda_c=2\pi c/\omega$,
\begin{eqnarray}
i\frac{\partial}{\partial t} \psi({\bf r},t) & = & 
\left[-\frac{1}{2}\nabla^2+V_{\rm eff}(r) + z \, F(t) \right] \psi({\bf r},t),
\label{eq:tdse}
\end{eqnarray}
where $F(t)=F_0 f(t) \sin(\omega t)$ denotes the laser electric field, $f(t)$ is the envelope function and $V_{\rm eff}(r)$ the atomic potential.
For hydrogen (H), $V_{\rm eff}(r)$ is the bare Coulomb potential while for argon (Ar) we employ a model potential \cite{ar:muller98} within the single-active electron approximation which reproduces the binding energy to an accuracy of typically $ \approx 10^{-3}$.
We employ two complementary methods to solve Eq.\ (\ref{eq:tdse}) in order to establish reliable and consistent results. 

In the first method, Eq.\ (\ref{eq:tdse}) is numerically integrated using the alternating direction
implicit (Peaceman-Rachford) method \cite{Kulander1992} with a uniform grid spacing $\Delta r$ being dependent on the numerical problem in the range of $10^{-2}\le \Delta r \le 6.25\times 10^{-2}$ a.u. In general, a finer grid spacing is needed for a longer wavelength, and also for Ar than for H.
In order to reduce the difference between the discretized and analytical wave function, we scale the
Coulomb potential by a few percent at the first grid point \cite{Krause1992}.
The time step $\Delta t$ is typically 1/16000 of an optical cycle for 800 nm wavelength, i.e., $6.895\times 10^{-3} {\rm a.u.}$. 
This algorithm is accurate to the order of $\mathcal{O}(\Delta t^3)$. 
In the second method,  the TDSE is integrated on a finite grid by means of the pseudo-spectral method \cite{ar:tong97} which is also accurate to the order of $\mathcal{O}(\Delta t^3)$. 
It allows for larger time steps of the order of 0.1 atomic units.
The $r$ coordinate is discretized within the interval $[0,r_{\rm max}]$ with a non-uniform mesh point distribution. 
The innermost grid point is typically as small as $2.5\times 10^{-4}$ a.u., enabling an accurate description near the nucleus.
A smooth cut-off function is multiplied at each time-step to avoid spurious reflections at the border $r_{\rm max}$, 
while another cut-off function prevents reflections at the largest resolved energy  $E_{\rm max}$. 
For Ar the occupied states supported by the model potential are dynamically blocked during the time evolution by assigning a phase corresponding to an unphysically large and positive energy eigenvalue \cite{ar:klaus_hhg06}.
We calculate the dipole acceleration $\ddot d(t) = -  \partial_t^2 \langle z(t) \rangle $, 
employing the Ehrenfest theorem through the relation $\ddot d(t)= \langle \psi({\bf r},t)\mid \cos \theta /r^2  - F(t) \mid \psi({\bf r},t)\rangle$\cite{ar:tong97}, where the second term can be dropped as it does not contribute to the HHG spectrum.

For the wavelength-dependence of the harmonic yield, in particular the global scaling, it is important to specify the definition  of the integral yield. One can focus on a given number of harmonic orders, on a fixed energy interval, or the entire spectrum. 
Following Refs.\ \cite{ar:tate_scaling07,Schiessl2007PRL,Frolov2008PRL,Schiessl2008JMO}, we consider in this work the HHG yield defined as energy radiated from the target atom (single-atom response) per unit time \cite{Jackson} integrated for a fixed photon energy range, specifically from 20 to 50 eV:
\begin{equation}
\Delta Y  =  \frac{1}{3 c^3T} \int_{20\,{\rm eV}}^{50\,{\rm eV}}  | a(\omega)|^2 d \omega,
\label{eq:yield}
\end{equation}
where $T$ denotes the pulse duration. Note that the energy window $\Delta E$ of the output radiation (here 20 to 50 eV) is kept constant when analyzing $\Delta Y$ as a function of $\lambda$. 
Clearly, both the number and order of the harmonic peaks lying in the fixed energy interval change as $\lambda_c$ is varied. 

\section{Wavelength dependence}
\label{sec:wavelength}

We adopt the laser parameters of Ref.\ \cite{ar:tate_scaling07}, with a fixed peak intensity of $1.6 \times 10^{14} $ W/cm$^2$, a variation of $\lambda$ between 800 nm and 2 $\mu$m, and an envelope function $f(t) $ corresponding to an 8-cycle flat-top sine pulse with a half-cycle turn-on and turn-off. 

\begin{figure}
\epsfig{file=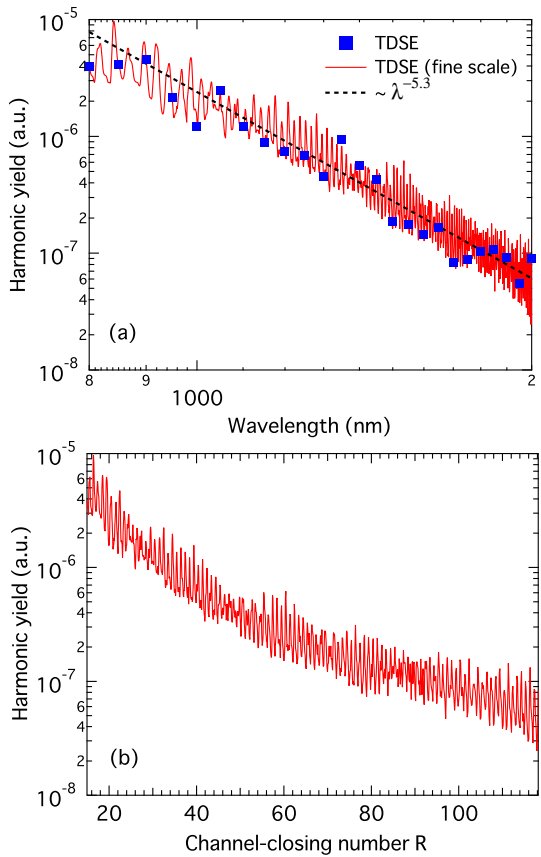,width=8.3cm,clip=} 
\caption{(Color online) Wavelength dependence of the integrated harmonic yield $\Delta Y$ between 20 and 50 eV as a function  of (a) wavelength $\lambda$ and (b) channel-closing number $R$. 
$\blacksquare$ : TDSE results obtained by the pseudo-spectral method.  
In addition, results on a fine scale (solid line) are presented. 
Dashed line: fit $\Delta Y \propto \lambda^{-5.3}$.}
\label{fig:H_800_2000}
\end{figure}

Figure \ref{fig:H_800_2000} displays the HHG yield for atomic hydrogen calculated on a fine mesh in $\lambda$ with a spacing of 1 nm. Superimposed on a global power-law dependence $\Delta Y\propto \lambda^{-x}$ ($x\approx 5$) \cite{ar:tate_scaling07,Schiessl2007PRL,Frolov2008PRL,Schiessl2008JMO}, we find remarkably strong and rapid fluctuations through the entire $\lambda$ range. The origin of this oscillation can be identified as the quantum interference of up to ten rescattering trajectories, based on the SFA analysis \cite{Schiessl2007PRL,Schiessl2008JMO}. 

The effect of the interference of multiple quantum paths was previously studied in the context of the \emph{intensity dependence} of HHG and above-threshold ionization (ATI) \cite{Borca2002PRA,Milosevic2002PRA,Zair2008PRL}. 
Using the quasistationary quasienergy state theory for a zero-range potential and the strong-field approximation, Borca {\it et al.} \cite{Borca2002PRA}, and Milo\v{s}evi\'c and Becker \cite{Milosevic2002PRA} have shown that HHG exhibits resonance-like enhancement when $N$-photon ionization channel is closed with increasing intensity, i.e. the parameter $R$ (Eq.\ (\ref{eq:cc})) becomes an integer. 
Za\"{\i}r \emph{et al.} \cite{Zair2008PRL} have very recently reported experimental observation of the oscillation in the intensity dependence of the HHG yield as evidence of the interference between the short and long paths. 
The analysis in Ref.\ \cite{Milosevic2002PRA} may be applied to the wavelength-dependence as well, suggesting to look at our results in terms of $R$. 
Data of Fig.\ \ref{fig:H_800_2000}(a) are replotted in terms of $R$ in Fig.\ \ref{fig:H_800_2000}(b), which permits a detailed analysis of the channel closing behavior (see below). 

In Figs.\ \ref{fig:H1000} and \ref{fig:Ar1950} we reexamine the role of quantum paths in the oscillations of the wavelength-dependence of the harmonic yield as a function of $R$ on a finer $R$ scale. 
We compare full TDSE solutions with approximations based on the strong field approximation (SFA) \cite{ar:lewenstein94,ar:ivanov96}. 
We first apply the Gaussian model \cite{ar:lewenstein94}, in which the ground-state wave function has the form
\begin{equation}
\label{eq:Gaussian_model}
\psi ({\bf r}) = \left(\frac{\alpha}{\pi}\right)^{3/4} e^{-\alpha {\bf r}^2/2},
\end{equation}
where $\alpha$ is chosen to reproduce $I_p$. An appealing point of the Gaussian model is that the dipole transition matrix element also takes a Gaussian form \cite{ar:lewenstein94},
\begin{equation}
\label{eq:GaussianDME}
{\bf d}({\bf p})=i\left(\frac{1}{\pi\alpha}\right)^{3/4}\frac{\bf p}{\alpha} e^{-{\bf p}^2/2\alpha},
\end{equation}
and that one can evaluate the integral with respect to momentum in the formula for the dipole moment (Eq.\ (8) of Ref.\ \cite{ar:lewenstein94}) analytically, without explicitly invoking the notion of quantum paths. 
Unphysically rapid decrease for ${\bf p}^2/2\alpha \gg 1$ limits the application of Eq.\ (\ref{eq:GaussianDME}) to harmonic orders with momenta of the returning electron not substantially exceeding ${\bf p}^2/2\alpha \approx 1$.  
We have confirmed that the resulting harmonic spectra have an adequate plateau and cut-off structure for the value of $\alpha$ (1 and 2 a.u.) used in the present study. The obtained wavelength dependence of the HHG yield, expressed in terms of $R$ (Fig.\ \ref{fig:H1000} (b)), exhibits oscillations similar to that in the TDSE result (Fig.\ \ref{fig:H1000} (a)), although peaks are found - contrary to TDSE results - near integer values of $R$. 
In addition we employ complex solutions of the saddle-point approximation (SPA) \cite{Milosevic2002PRA}, while we have previously obtained similar results by employing classical trajectories \cite{Schiessl2007PRL,Schiessl2008JMO}.  
Up to 16 possible trajectories for each individual photon energy are considered.
When including up to ten and twelve returning paths for the case of H and Ar, respectively, the SPA can reproduce the modulation depth and frequency of the $\lambda$ oscillations of the TDSE and the Gaussian model reasonably well, thus strongly supporting the quantum path interference as the origin of the fluctuations. The SPA result for Ar with twelve trajectories (Fig.\ \ref{fig:Ar1950} (b)) reproduces even the small peaks between the main peaks.

\begin{figure}
\centerline{
\epsfig{file=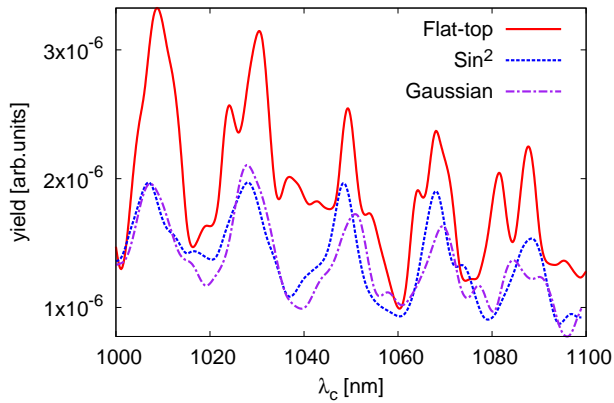,width=8.3cm,clip=} 
}
\caption{(Color online) Fluctuations of the harmonic yield $\Delta Y$ as a function of the fundamental wavelength $\lambda$  for hydrogen.
Solid: 8-cycle flat-top with a $1/2$ ($1/2$) cycle ramp on (off), 
dotted: 14-cycle $\sin^2$ pulse with a FWHM  of $\tau_p=7$ cycles, 
dash-dotted: 16-cycle Gaussian with a FWHM of $\tau_p=7$ cycles.
Other pulse parameter are the same as in Fig.\ \ref{fig:H_800_2000}. 
}
\label{fig:pulsedependence}
\end{figure}

We emphasize the remarkable variation on a fine $\lambda$ scale. One might suspect that the oscillation as in Figs.\ \ref{fig:H_800_2000}--\ref{fig:Ar1950} would be specific to monochromatic driver pulses and smeared out for the case of ultrashort broadband pulses. The pulse shape used in this study is, however, not monochromatic but its spectral width $\Delta \lambda$ is $\sim 10$\% of $\lambda_c$. 
The rapid variations of the harmonic yield occur on a scale $\delta \lambda$ much smaller than this width. This finding, at a first glance surprising, is a direct consequence of the quantum path interference. It follows from the existence and the fixed spacing in between discrete points in time - controlled by $\lambda_c$ -  at which electronic trajectories are launched. As long as the few-cycle pulse permits the generation of a set of a few quantum paths in subsequent half-cycles, the overall temporal characteristics of the driver pulse is of minor importance. We have also checked that the fluctuations in the harmonic yield are not an artefact of our particular choice of $f(t)$. They can be observed also for ``smoother'' pulse shape such as $\sin^2$ and Gaussian pulses as well as shorter pulses, provided that the pulse can support multiple returning trajectories (Fig.\ \ref{fig:pulsedependence}). 
The temporal profile of the pulse influences the detailed shape of the interference pattern, 
in particular the amplitude of the oscillations decreases with decreasing pulse length due to the reduction of the effective number of returning electron trajectories. 

\begin{figure}
\epsfig{file=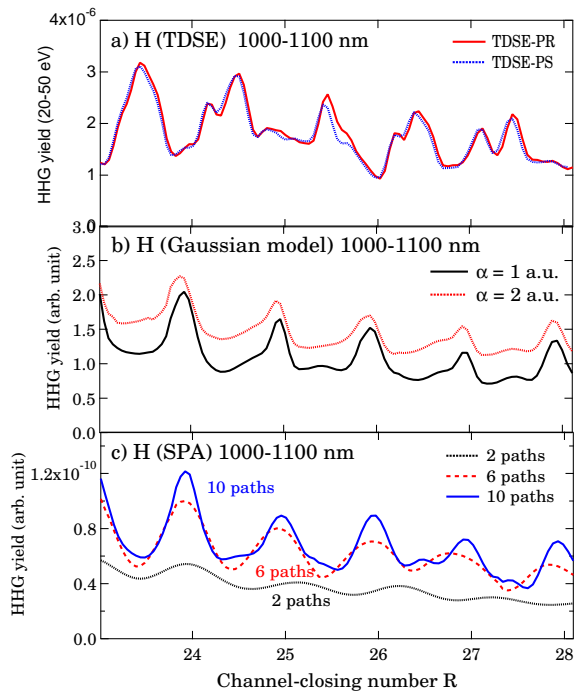,width=8.3cm,clip=}
\caption{(Color online) 
Variations of the integrated harmonic yield (20 to 50 eV) in a narrow range of $\lambda=1000-1100$ nm, as a function of $R$, for H. 
a) comparison between the TDSE solutions with the Peaceman-Rachford (PR) and the pseudo-spectral (PS) methods.
b) the results of the Gaussian model with $\alpha=1\,{\rm a.u.}$ (solid) and 2 a.u. (dotted). 
c) build-up of the interference pattern with increasing number of quantum trajectories within the SPA. 
In b) and c) the vertical axis is in arbitrary units.
\label{fig:H1000}}
\end{figure}

\begin{figure}
\epsfig{file=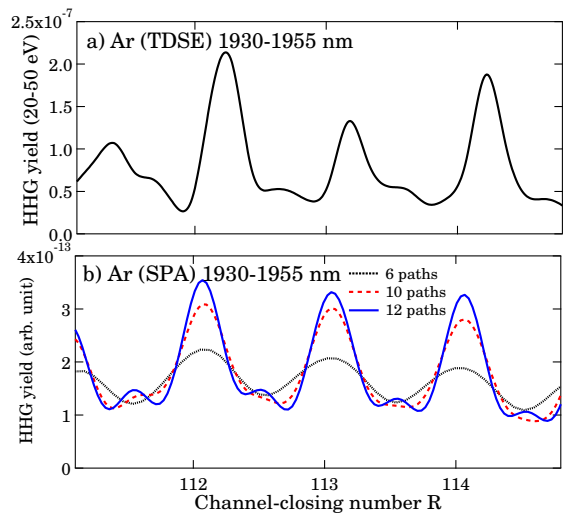,width=8.3cm,clip=}
\caption{(Color online) Variations of the integrated harmonic yield (20 to 50 eV) in a narrow range of $\lambda=1930-1955$ nm, as a function of $R$, for Ar.
a) the TDSE solution,
b) build-up of the interference pattern with increasing number of quantum trajectories within the SPA.
\label{fig:Ar1950}}
\end{figure}

\section{Intensity dependence}
\label{sec:intensity_dependence}
\begin{figure}
\centerline{
\epsfig{file=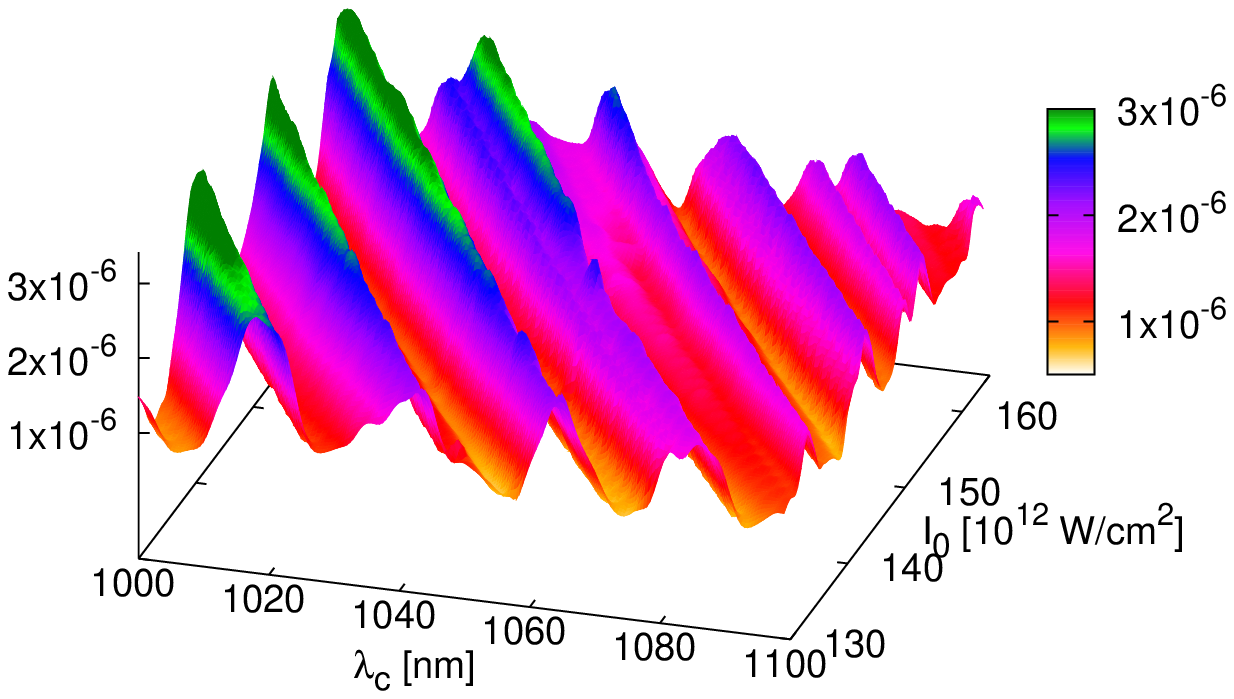,width=8.cm,clip=} 
}
\centerline{
\epsfig{file=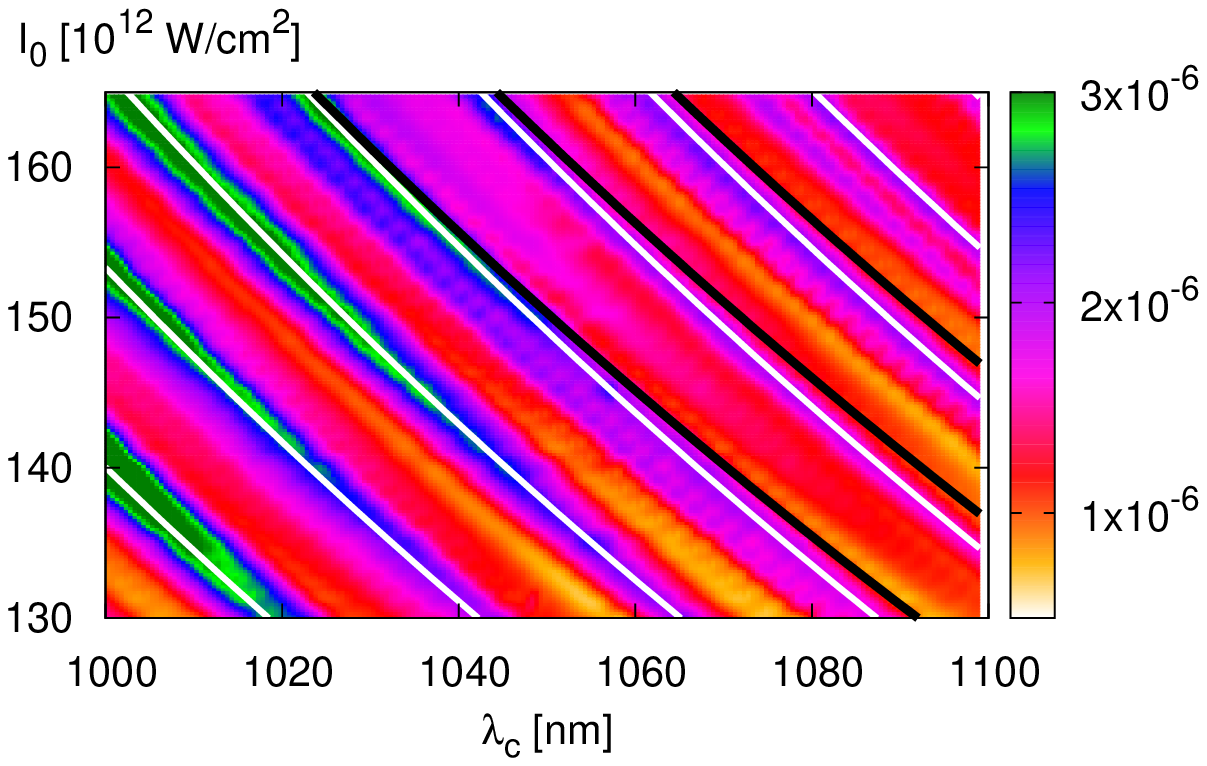,width=8.cm,clip=} 
}
\caption{(Color online) TDSE-calculated integrated harmonic yield between 20 and 50 eV for H (8-cycle flat-top pulse) in the ($\lambda_c,I_0$) plane. 
In the contour plot (lower panel), white lines show values of constant $(U_p+I_p)/\hbar\omega$, shifted from integer values by +0.52, while black lines (only three are shown for clarity) represent values of constant $(U_p+\tilde I_p)/\hbar\omega$ with $\tilde I_p=10.5$ eV. 
}
\label{fig:tdse_cc_intensity_lambda}
\end{figure}

\begin{figure}
\centerline{
\epsfig{file=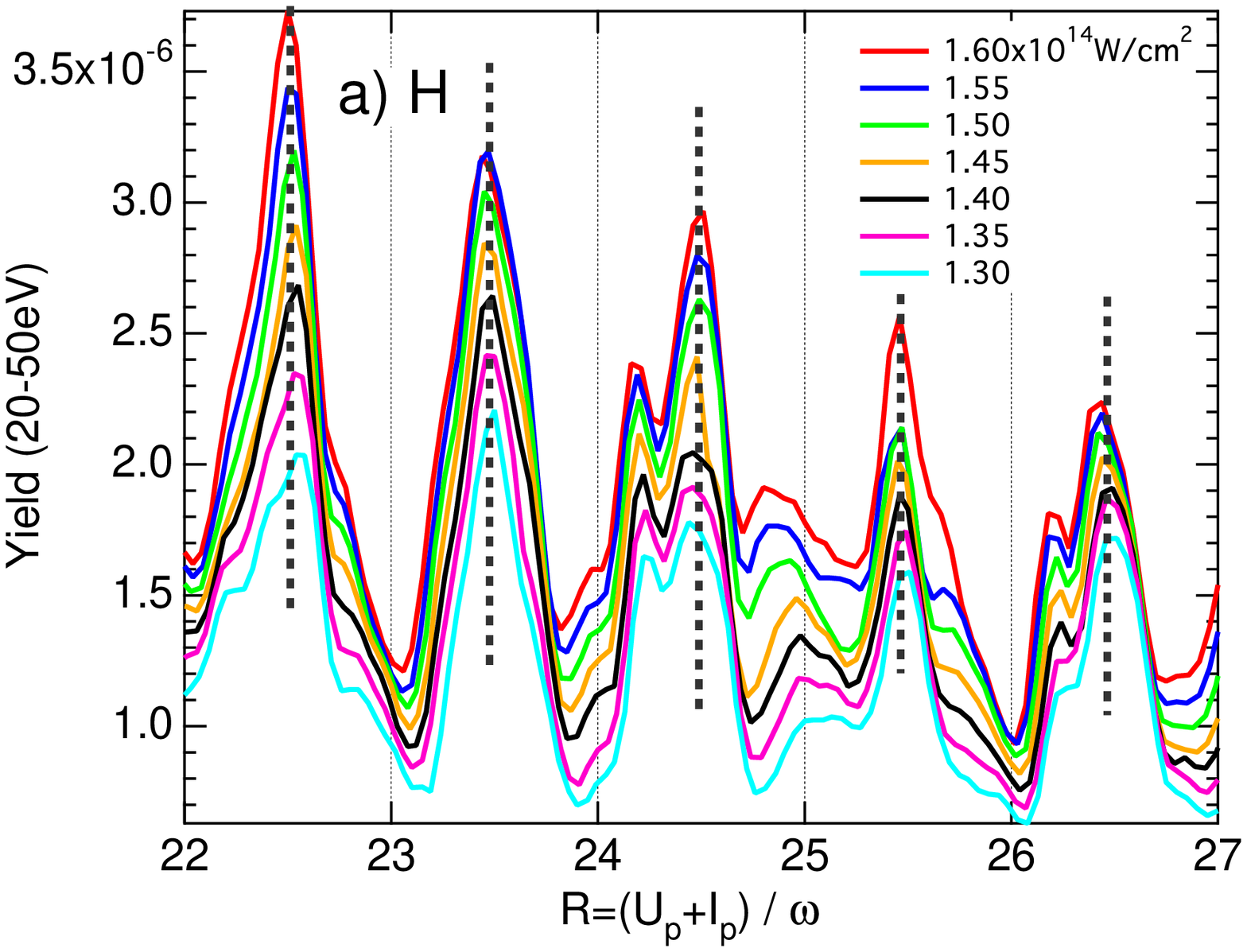,width=8.3cm,clip=} 
}
\centerline{
\epsfig{file=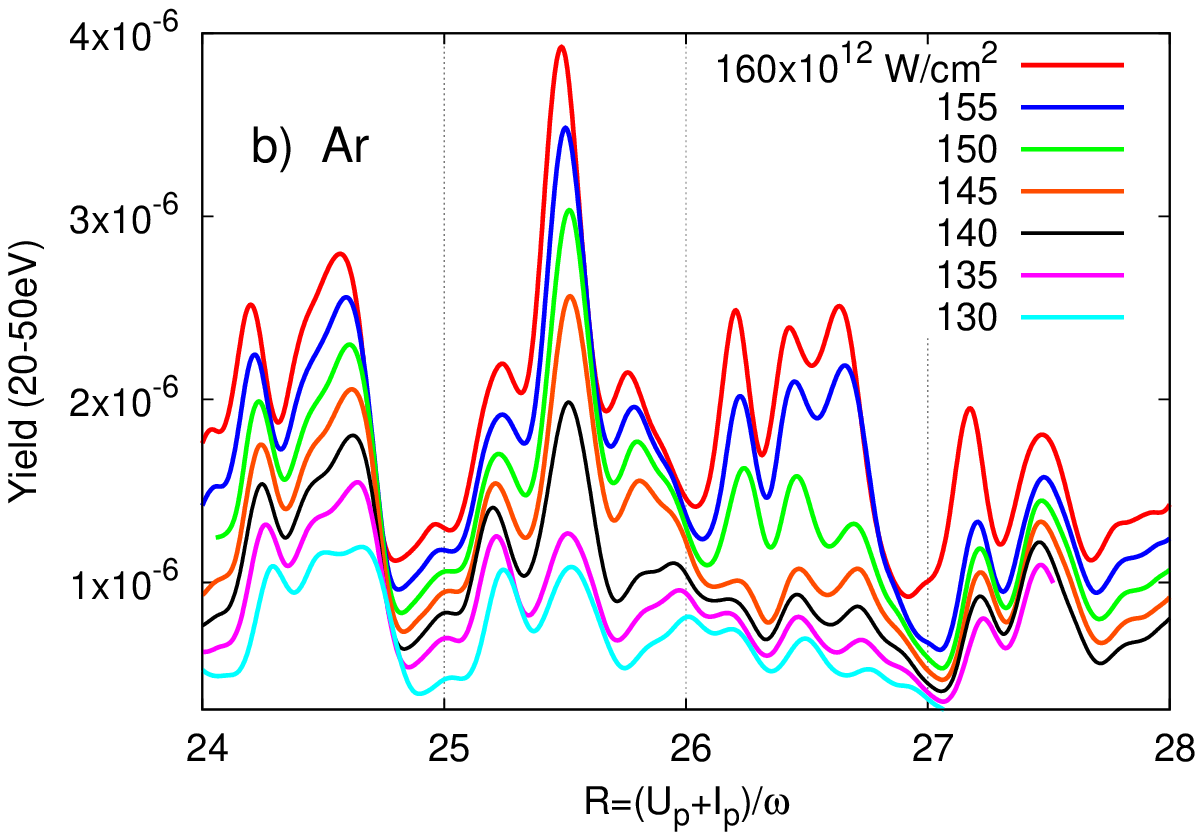,width=8.3cm,clip=} 
}
\caption{(Color online) 
Wavelength dependence of the integrated harmonic yield (20 to 50 eV) in the range of $\lambda\approx 1000-1100$ nm, expressed in terms of $R$, for a) H and b) Ar, for 8-cycle flat-top pulses for different intensities indicated in the figure.
}
\label{fig:dual_H_and_Ar}
\end{figure}

Previous work \cite{Zair2008PRL,Borca2002PRA,Milosevic2002PRA} studied the intensity dependence of the HHG yield at a fixed value of fundamental wavelength $\lambda_c$. 
On the other hand, we have so far focussed on the wavelength dependence at a fixed value of intensity ($1.6\times 10^{14} {\rm W/cm}^2$). 
We extend now this analysis to the two-dimensional parameter plane ($\lambda_c,I_0$) in order to explore the underlying mechanisms in more detail. 
An example for hydrogen (Fig.\ \ref{fig:tdse_cc_intensity_lambda}) for a narrow interval of wavelength (1 $\mu{\rm m} \le \lambda_c \le 1.1 \mu{\rm m}$) and intensity ($1.3\times 10^{14} {\rm W/cm}^2 \le I_0 \le 1.6\times 10^{14} {\rm W/cm}^2$) displays regularly shaped ridges each of which can be mapped onto a fixed channel closing number $R$. 
This regularity is also reflected in the cuts through this two-dimensional data for different fixed intensities for both hydrogen ( Fig.\ \ref{fig:dual_H_and_Ar} (a)) and argon (Fig.\ \ref{fig:dual_H_and_Ar} (b)).

Not only the peak positions but also detailed structures of the dependence on $\lambda_c$ are quite robust against the variation of $I_0$, when expressed in terms of $R$. 
For later reference we stress that these remarkable observations hold true only when the channel-closing (CC) number $R$ is determined with the \emph{true ionization potential} (Eq.\ (\ref{eq:cc})); the use of any other value of effective ionization potential would shift each peak and consequently each curve in Fig.\ \ref{fig:dual_H_and_Ar} by a different amount. 
This can also be understood from the fact that lines of constant values of $(U_p+\tilde I_p)/\hbar\omega$ (with e.g.\ $\tilde I_p=10.5$ eV) in Fig.\ \ref{fig:tdse_cc_intensity_lambda} (b) deviate from the ridges which manifest as peaks along lines of constant $I_0$ (Fig.\ \ref{fig:dual_H_and_Ar} (a)).   
Results for argon (Fig.\ \ref{fig:dual_H_and_Ar} (b)) show a similar behavior, indicating the applicability of the parameter $R$ independent of the atomic species.

\section{Modulation period}
\label{sec:modulation_period}

\begin{figure}
\epsfig{file=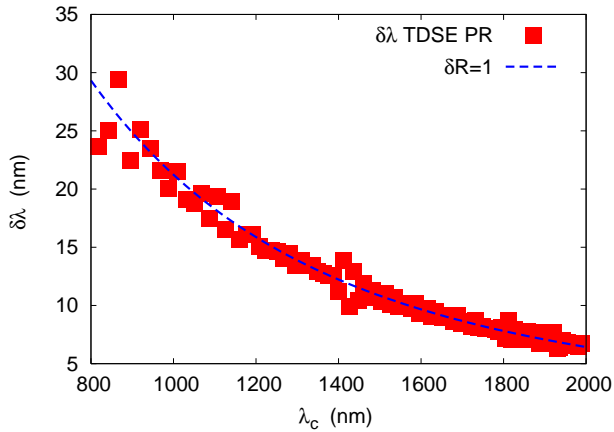,width=8.3cm,clip=}
\caption{(Color online) Variation of the modulation period $\delta\lambda$ with the driver central wavelength $\lambda_c$ for atomic hydrogen. \Red{$\blacksquare$}: TDSE, dashed line: $\delta R=1$ (Eq.\ (\ref{eq:formula_deltalambda1})).}
\label{fig:deltalambda}
\end{figure}

The modulation period $\delta \lambda$ of the harmonic yield is a function of the central wavelength $\lambda_c$ itself. With increasing $\lambda_c$, $\delta \lambda$ decreases from about 30 nm near 800 nm wavelength to $\approx 6$ nm near a wavelength of 2 $\mu$m (Fig.\ \ref{fig:deltalambda}). 
However, expressed in terms of the channel closing number $R$, the separation of the principal peaks is given by $\delta R=1$ both for the TDSE and the SPA results (see Figs.\ \ref{fig:H1000} and \ref{fig:Ar1950}). 
Interference peaks appear with this spacing regardless of intensity (Fig.\ \ref{fig:dual_H_and_Ar}). 
The peaks in the TDSE results are, however, \emph{not} located at integer values of $R$, as opposed to the SPA results as well as previous theoretical work \cite{Borca2002PRA,Milosevic2002PRA}. This problem was previously encountered in the intensity-dependence of HHG and ATI \cite{Kopold2002JPB}. 
In order to recover integer values the use of an effective ionization potential $\tilde{I}_p$ in place of $I_p$ in Eq.\ (\ref{eq:cc}) was proposed based on arguments that either the enhancement was due to multiphoton resonances with ponderomotively upshifted Rydberg states \cite{Muller1999PRA} ($\tilde{I}_p$ corresponds to the excitation energy of the resonant state) or that high-lying atomic states are strongly distorted by an intense laser field to form a quasicontinuum, effectively lowering the ionization potential \cite{Frolov2008PRL}. 
As long as one considers only the intensity-dependence at a fixed wavelength, the difference $\Delta \tilde{I}_p=\tilde{I}_p-I_p$ causes a constant shift of $R$ by $\Delta \tilde{I}_p/\omega$. 
Consequently, integer values of $R$ could be restored along this axis for a suitable choice of $\tilde{I}_p$. 
However, considering now the wavelength-dependence at a fixed intensity, $\Delta \tilde{I}_p/\omega$ itself would depend on $\lambda$. Therefore, if the modulation period $\delta\lambda$ corresponds to $\delta R = 1$ for the true $I_p$, any other choice of $\tilde{I}_p$ different from $I_p$ cannot shift all the peaks uniformly to integer values of $R$. 

\begin{figure}
\epsfig{file=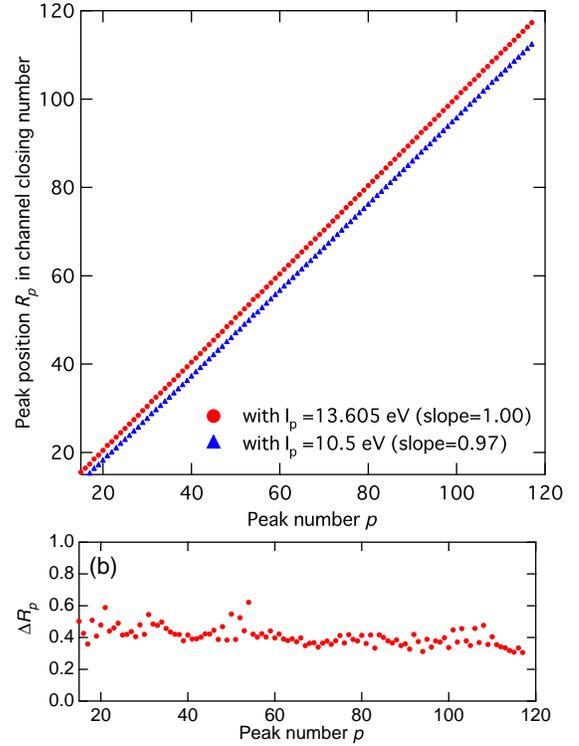,width=8.3cm,clip=}
\caption{(Color online) 
Position $R_p$ of the principal peaks ($p=15-117$) from Fig.\ \ref{fig:H_800_2000}(b). \Red{$\bullet$}: $(U_p+I_p)/\hbar\omega$ with the true $I_p$ = 0.5 a.u.\,(13.605 eV), \Blue{$\blacktriangle$}: $(U_p+\tilde{I}_p)/\hbar\omega$ with $\tilde{I}_p$ = 10.5 eV. The slope obtained by line fitting is also indicated. 
(b) Corresponding $\Delta R_p$ values (Eq.\ (\ref{eq:peakshift})) with the true $I_p$ = 0.5 a.u..
}
\label{fig:H_peaks}
\end{figure}

In a further step, we enumerate all the principal peaks in Fig.\ \ref{fig:H_800_2000} (b) from $p=15$ to 117, and plot the CC number $R_p$ as well as the mismatch to the nearest integer,
\begin{equation}
\label{eq:peakshift}
\Delta R_p\equiv R_p-[R_p],
\end{equation}
of each peak in Fig.\ \ref{fig:H_peaks}. 
The slope of the line fitted to the data calculated with true $I_p$ (filled circles in Fig.\ \ref{fig:H_peaks}(a)) is nearly equal to unity ($\approx 1.00$), while those with $\tilde{I}_p=10.5\,{\rm eV}$ (diamonds in Fig.\ \ref{fig:H_peaks}(a)) have a slope smaller than unity. 
Moreover, although some fluctuation is seen, the values of $\Delta R_p$ are roughly constant, most of them being distributed between 0.3 and 0.6.  

As can be seen from Figs.\ \ref{fig:H1000}--\ref{fig:dual_H_and_Ar}, the harmonic yields $\Delta Y$ are not only composed of peaks separated by $\delta R=1$, but also often contain finer structures with sub-peaks. 
This is even more so for longer pulses. 
In order to extract the periodicity of these structures quantitatively we calculate the power spectrum of $\Delta Y(R) \times \lambda^{5.3}$ where the multiplication by $\lambda^{5.3}$ removes the smooth global decay (Fig.\ \ref{fig:H_800_2000} (b)). 
We can clearly identify the sharp dominant frequency component precisely at $\Omega=1$, corresponding to $\delta R=1$. A beat-like structure of a period of $\approx 20$ seen in Fig.\ \ref{fig:H_800_2000} (b) gives rise to an additional small side band.

The Fourier spectrum clearly underscores that the peak separation corresponds to $\delta R=1$ throughout the entire wavelength range between 800 nm and 2 $\mu$m.
\begin{figure}
\epsfig{file=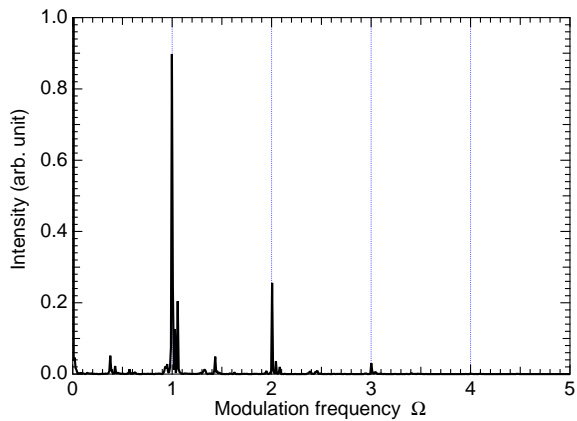,width=8.3cm,clip=}
\caption{Power spectrum of $\Delta Y(R) \times \lambda^{5.3}$ (see Fig.\ \ref{fig:H_800_2000} (b)).}
\label{fig:H_peaks_FT}
\end{figure}
This spacing is closely related to the spacing $\delta \lambda$ in the wavelength dependence of the peak positions. Hence, we can derive the scaling of $\delta \lambda$ with $\lambda$ to obtain,
\begin{eqnarray}
\label{eq:formula_deltalambda1}
\delta\lambda&=&\frac{const.}{I_p+3U_p},\nonumber\\
&=& \frac{1240}{I_p({\rm eV})+2.8\times 10^{-19}I({\rm W/cm}^2)\lambda^2({\rm nm})}\,{\rm nm}.
\end{eqnarray}
Equation \ref{eq:formula_deltalambda1} reproduces the TDSE-calculated $\lambda$ dependence of $\delta\lambda$ quite well (Fig. \ref{fig:deltalambda}).

The present results as well as those in Sec.\ \ref{sec:intensity_dependence} strongly indicate that the wavelength and intensity dependence of the HHG yield calls for an explanation in terms of $R$ calculated from the true $I_p$ in spite of the pronounced shift of the peak position from integer $R$ values.\\[0.5cm]

\section{Peak shift from integer $R$ values}
\label{sec:truncated_Coulomb}

While the peak separation is given by $\delta R=1$, enhancements do not appear at $R=N$, with $N$ being an integer but shifted by an amount ranging from 0.3 to 0.6 (see Fig.\ \ref{fig:H_peaks} (b)). 
This is in clear contrast to the SFA prediction for the CC peaks in the literature \cite{Milosevic2002PRA} as well as to our present SFA results in Figs.\ \ref{fig:H1000} (b)(c) and \ref{fig:Ar1950} (b). 
The fundamental difference between the SFA and the full solution of the TDSE is - apart from numerical or analytical solution strategies - 
that, in the former one neglects the excited states and the effect of the atomic potential to continuum electrons, 
which may be a serious deficiency for long-range potentials such as the Coulomb potential. 
Recent studies on ionization dynamics and doubly-differential photoelectron momentum distributions of hydrogen have shown the significance of the long-ranged Coulomb potential in laser-atom interaction and have illustrated the failure of the SFA near the threshold \cite{ar:diego_prl06,ar:diego_pra07}.

\subsection{Truncated Coulomb potential}
\label{se:truncated}

In order to explore the significance of the long tail of the Coulomb potential for the present case of interferences in the HHG yield, 
we perform calculations with a truncated Coulomb potential, given by
\begin{equation}
\label{eq:truncated_potential}
V_{eff}(r,r_c) = \left\{\begin{array}{ll}-\frac{1}{r} & (r < r_c) \\ -\frac{e^{-(r-r_c)/r_d}}{r} & (r > r_c)\end{array}\right.\qquad ,
\end{equation}
where the effective range of the truncated Coulomb potential $r_c$ is varied between $r_c=10$ and $r_c=70$ a.u.
and the width of the cross-over region $r_d$ is chosen to be $r_d=10$ a.u.. 
For these parameter values, the ionization potential and the first excitation energy remain unchanged to an accuracy of $\approx 10^{-9}$ and $\approx 10^{-3}$, respectively. It should be noted that the classical electron quiver motion amplitude is $\alpha_q=26.3$ a.u.\ for $I=1.6\times 10^{14} {\rm W/cm}^2$ and $\lambda_c = 900$ nm, and $\alpha_q=39.3$ a.u.\ for $\lambda_c = 1100$ nm. 

\begin{figure}
\centerline{\epsfig{file=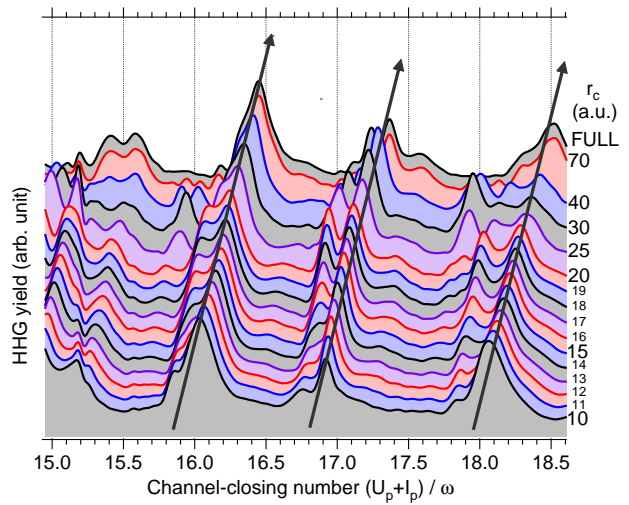,width=8.3cm,clip=} }
\caption{(Color online) Comparison of the wavelength-dependence $800\,{\rm nm}< \lambda_c <900\,{\rm nm}$ of the harmonic yield $\Delta Y$ for H, expressed in terms of $R$, calculated with the full Coulomb potential (marked as ``FULL") with the truncated Coulomb potentials for varying values of $r_c$ as indicated. 
The pulse has a 16-cycle flat-top shape, other pulse parameters are the same as in Fig.\ \ref{fig:H_800_2000}. }
\label{fi:details_hydrogen_40scut}
\end{figure}

\begin{figure}
\centerline{\epsfig{file=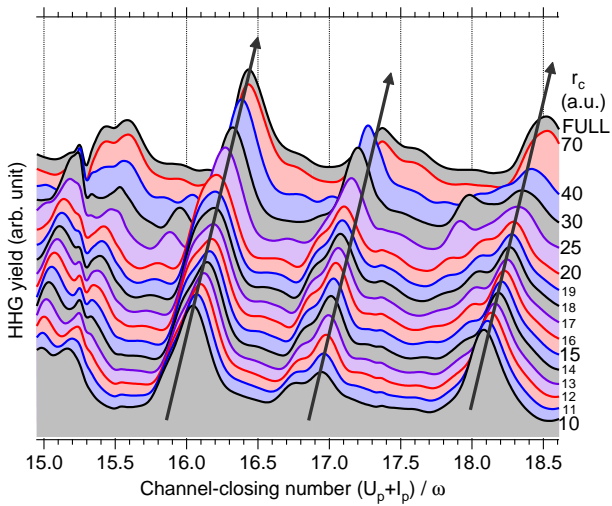,width=8.3cm,clip=} } 
\caption{(Color online)
Same as Fig.\ \ref{fi:details_hydrogen_40scut}, but for a pulse length of 8 cycles (flat-top). 
}
\label{fi:details_hydrogen_40scut_8cyc}
\end{figure}

We thus explore the entire range from $r_c/\alpha_q \approx 0.3$ to $r_c/\alpha_q \approx 2.3$. 
Convergence to the solution employing the full Coulomb potential is reached only for $r_c$ as large as 70 a.u.\ (see Figs.\ \ref{fi:details_hydrogen_40scut} and \ref{fi:details_hydrogen_40scut_8cyc}). 
Most important in the present context is a systematic, almost rigid shift of the peaks as a function of $r_c$. 
Only for small $r_c$ ($\approx$10 a.u.), the maxima are found near channel closings (near $R$ equal to an integer), in agreement with the SFA results \cite{Milosevic2002PRA}.
This observation indicates that the Coulomb potential is indeed responsible for a - to first approximation - monotonic and nearly uniform shift of the peaks. 
It should be noted that the long-ranged Coulomb potential manifests itself in two seemingly different effects. 
Firstly, it supports high-lying Rydberg states which converge to the continuum at threshold. 
Furthermore, Coulomb scattering and deflection influences the motion of the returning electron, even at large distances from the core.

\subsection{Effective ionization potential}
\label{se:effectiveip_from_tdse}

The influence of the potential form on the position of the CC was previously identified within the framework of a 1D-TDSE model \cite{ar:faria02_closings,ar:taieb03_closings}.
It was suggested to use an ``effective" ionization potential $\tilde I_p$ in Eq.\ (\ref{eq:cc}) when comparing TDSE calculation with models employing zero-range potentials to account for high-order above-threshold
ionization (ATI) spectra at $R \ne N$ \cite{Kopold2002JPB}. 
Employing an effective ionization potential $\tilde I_p$ in an SFA model roughly leads to a rigid horizontal shift of the interference structure of the HHG yield, in accordance with our observations in Figs.\ \ref{fi:details_hydrogen_40scut} and \ref{fi:details_hydrogen_40scut_8cyc}. 

Different lines of arguments are invoked for employing $\tilde I_p$ rather than $I_p$.  
However, they all have in common that the existence of a strongly distorted, continuum-like excited state $\varepsilon_n$ is considered responsible for an effectively {\it lower} ionization threshold. 
For convenience, let us therefore define $\Delta \tilde I_p \equiv \tilde I_p -I_p$, which is expected to be a negative quantity ($\Delta \tilde I_p < 0$). 
Different choices of $\Delta \tilde I_p$ are explored. 
Faria {\it et al.} \cite{ar:faria02_closings} argue that $\varepsilon_n$ should be given by the condition that its radius $r_n\approx 3n^2/2$ for principal quantum number $n$ should match the quiver amplitude $\alpha_q=F_0/\omega^2$. 
Together with the Rydberg energy $\varepsilon_n\approx -I_p/n^2$ this would imply
\begin{equation}
\label{eq:Faria}
\Delta \tilde I_p \approx -\frac{3\,\omega^2}{2F_0}I_p \propto I^{-1/2} \lambda^{-2}.
\end{equation}
Accordingly, the change of the effective ionization potential, $\Delta \tilde I_p$, becomes wavelength dependent. 
On the other hand, we have found a fairly rigid equidistancy $\delta R=1$ as well as a nearly constant shift $\Delta R$ of the latter away from the integers over a wide range of $\lambda_c$. 
Consequently, if we assume that $\Delta \tilde I_p$ compensates for $\Delta R_p$, these quantities must satisfy the relation $\Delta \tilde I_p = -(\Delta R_p+m)\hbar\omega$, with $m$ being a possible integer offset and $\Delta R_p$ defined in Eq.\ (\ref{eq:peakshift}). Figures \ref{fi:details_hydrogen_40scut} and \ref{fi:details_hydrogen_40scut_8cyc} clearly show that the amount of the peak shift in $R$ is smaller than unity, hence $m=0$. This leads to
\begin{equation}
\label{eq:condition}
\Delta \tilde I_p = -\Delta R_p\hbar\omega \propto \lambda^{-1},
\end{equation}
Obviously, hypothesis Eq.\ (\ref{eq:Faria}) is not consistent with Eq.\ (\ref{eq:condition}). 
In addition, no upper limit for $\Delta \tilde I_p$ according to Eq.\ (\ref{eq:Faria}) was discussed in literature.  
This may lead to the obviously incorrect conclusion that $\Delta \tilde I_p \to 3.4$ eV as soon as in a low intensity and low wavelength limit the $n=2$ Rydberg state (or even the ground state!) would govern the effective threshold invoked. 

An alternative proposal put forward by Frolov {\it et al.} \cite{Frolov2008PRL} relates the energy $\varepsilon_n$ to the formation of an effective continuum by broadening of the level with principal quantum number $n$. 
Accordingly, $\varepsilon_n$ is determined by the condition $\Gamma_n=\Delta \varepsilon_n$, where the width $\Gamma_n$ (related to ionization rate) approaches the level spacing $\Delta \varepsilon_n$. 
While in the limit of quasi-static tunneling, 
the tunneling rate $\Gamma_n \propto \exp{ \left[-2 (2 |\varepsilon_n|)^{3/2}/(3 F_0) \right]}$ strongly depends on the field strength $F_0$, but only very weakly on $\lambda$. 
The resulting value of $\Delta \tilde I_p$ is estimated to be -3.1 eV for atomic
hydrogen and $I=1.6\times10^{14}$ W/cm$^2$ in Ref.\ \cite{Frolov2008PRL}. This does not meet the condition Eq.\ (\ref{eq:condition}), according to which $|\Delta \tilde I_p|$ should be smaller than the photon energy $\hbar\omega$ (<1.5 eV in the present parameter range) of the driving laser pulse. 

\begin{figure}
\centerline{\epsfig{file=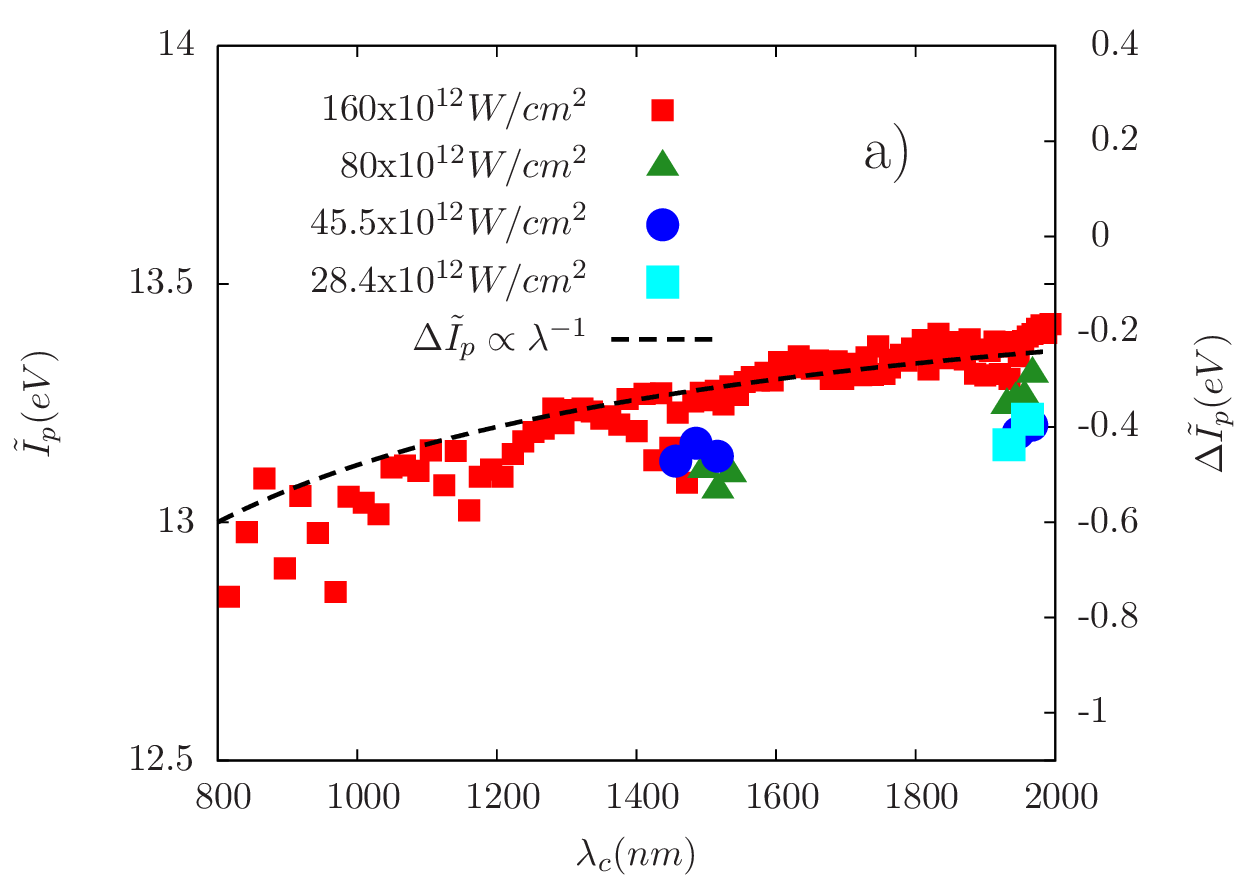,width=8.3cm}}
\centerline{\epsfig{file=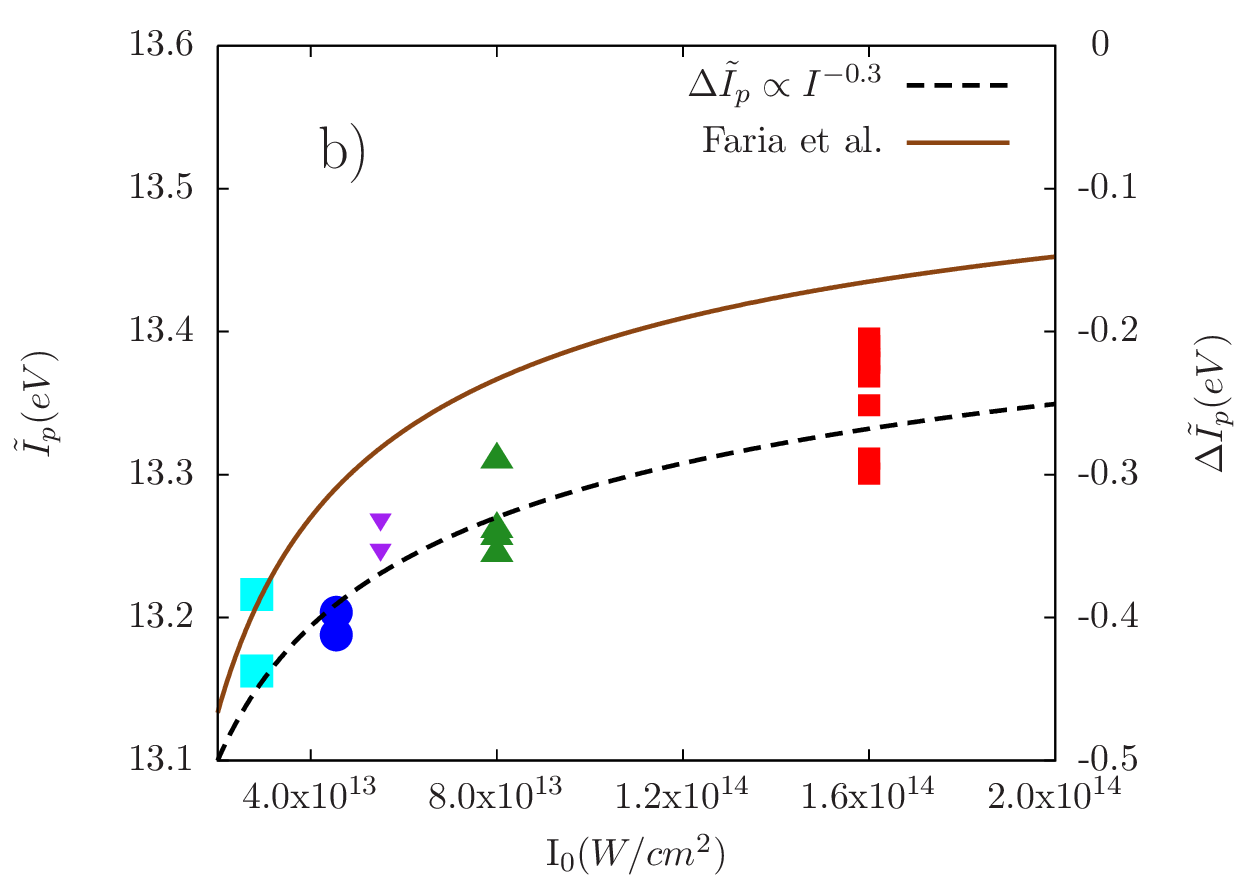,width=8.3cm}}
\caption{(Color online) $\tilde I_p$ (left axis) and $\Delta \tilde I_p$ (right axis) as obtained from Eq.\ (\ref{eq:condition}) for a broad range of driver wavelength and various intensities. 
{\bf a)} as function of $\lambda_c$, 
{\bf b)} as function of intensity near $\lambda_c = 1950$ nm. 
The solid line compares to the prediction of Faria {\it et al.} \cite{ar:faria02_closings}, 
while the dashed line shows a power law fit to our data ($\Delta \tilde I_p \propto I^{-0.3}$). 
}
\label{fi:hydrogen_ipeff_o}
\end{figure}

Moreover, with the help of Eq.\ (\ref{eq:condition}) we can determined the effective parameter dependence of $\Delta \tilde I_p$ employing the numerical values for $\Delta R$ over a wide range of $\lambda_c$ and $F_0$ (Fig.\ \ref{fi:hydrogen_ipeff_o}).
Our results suggest a weak dependence of $\tilde I_p$ on both the wavelength and the intensity, the latter being roughly proportional to $\lambda^{-0.3}$. 
This supports neither the explanations of Frolov {\it et al.} nor of Faria {\it et al.}. 

\subsection{Coulomb-corrected classical trajectory model}
\label{se:ccct}

In addition to the ability to support (an infinite number) excited bound states, the Coulomb potential affects the propagation of the rescattering electron which is neglected in the SFA as well. 
As the quantum interference of electron trajectories is responsible for the oscillation in the harmonic yield, their distortion by the potential may be crucial. 

In SFA, the time-dependent dipole moment $d(t)$ can be expressed as \cite{ar:ivanov96} 
\begin{equation}
d(t_f)= \sum_{P(t_i)} b_{\rm ion}(t_i) \cdot e^{ - i S_P(t_i,t_f,I_p) } \cdot c_{\rm rec}(t_f) + {\rm c.c.},
\label{eq:SFA_dipole}
\end{equation}
i.e., a sum over paths $P$ that start at the moment of tunnel ionization $t_i$ with amplitude $b_{\rm ion}(t_i)$, 
evolve in the laser field  - acquire the phase $e^{ - i S_P(t_i,t_f) }$ - and recombine upon rescattering at the core at time $t_f$ with the amplitude $c_{\rm rec}(t_f)$.
The interference oscillations in the HHG yield are controlled by the semiclassical action $S_P$ of the path $P$, which reads:
\begin{equation}
S_P(t_i,t_f,I_p) = \int_{t_i}^{t_f} \frac{(p + A(t'))^2}{2} dt' +I_p (t_f - t_i),
\label{eq:semiclassical_phase}
\end{equation}
where $A(t)$ is the laser vector potential, and $p$ is the classical momentum of the returning trajectory. 
The effect of the Coulomb potential can be incorporated into Eq.\ (\ref{eq:semiclassical_phase}) with help of an eikonal approximation as a correction to the (action) phase with \cite{ar:ivanov96},
\begin{equation}
\Delta S_P(t_i,t_f) = \int_{t_i}^{t_f}  \, V_{EI} (r(t^\prime))  \, dt' .
\label{eq:coulomb_correction}
\end{equation}
Clearly, the eikonal approximation would fail at small distances from the nucleus. 
This difficulty can by bypassed using the observation (Sec.\ \ref{se:truncated}) that at a cut-off $r_c=10$ a.u.\ the SFA limit of channel closings at integer values of $R$ is reached. 
Consequently, we set
\begin{equation}
V_{EI}(r)= V_{eff}(r,r_c=\infty ) - V_{eff}(r,r_c=10)
\label{eq:correction_potential}
\end{equation}
when calculating the long-range phase correction. 

We evaluate Eq.\ (\ref{eq:coulomb_correction}) along classical trajectories in the laser electric field $F(t)$, confined in the $xy$-plane, starting and ending at the ``tunnel exit" $z_0=I_p/F_0$. 
Trajectory modifications due to the Coulomb potential are small and can be neglected to first approximation \cite{ar:ivanov96}, i.e., we use the same sets of $(t_i,t_f)$ as in Eq.\ (\ref{eq:semiclassical_phase}). 

\begin{figure}
\centerline{\epsfig{file=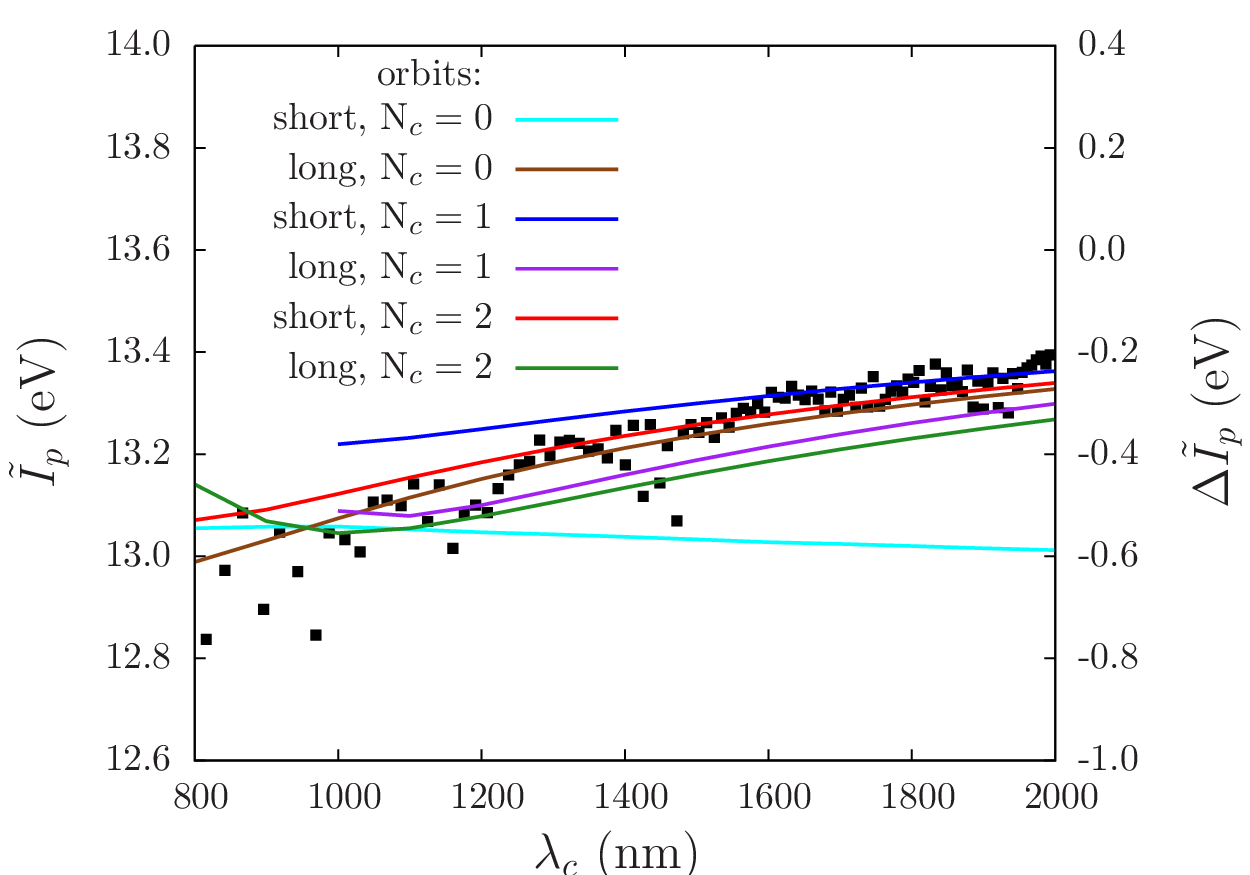,width=8.3cm,clip=}}
\caption{(Color online) 
$\tilde I_p$ (left axis) and $\Delta \tilde I_p$ (right axis) as a function of driver wavelength, employing Eq.\ \ref{eq:delta_Ip_corrected_coulomb} and the potential of Eq.\ \ref{eq:correction_potential}. 
Intensity is $I=1.6\times 10^{14}$ W/cm$^2$.
Lines stem from the six shortest orbits recolliding with 20eV, revisiting the core $N_c$ (here: zero, one, or two) times before recombining (as indicated). 
For each energy and each $N_c$, a short and a long orbit exist. 
$\blacksquare$: TDSE data (see also Fig.\ \ref{fi:hydrogen_ipeff_o}). 
}
\label{fi:ccct}
\end{figure}

It is now suggestive to express this additional phase in terms of a change in the ``effective ionization potential", $\tilde I_p$. Accordingly, 
\begin{equation}
\label{eq:delta_Ip_corrected_coulomb}
\Delta \tilde I_p= \Delta S_P(t_i,t_f)/(t_f-t_i).
\end{equation}
Figure \ref{fi:ccct} shows $\tilde I_p$ obtained by Eq.\ \ref{eq:delta_Ip_corrected_coulomb} for several (the shortest) classical trajectories which contribute to the harmonics near 33.6 eV (hence near the center of the HHG yield range considered in this work). 

Remarkably, most trajectories (save the shortest one) behave qualitatively very similarly.  
In spite of its oversimplification, this Coulomb-corrected model explains the behavior of $\tilde I_p$ even quantitatively well, which is a strong indication that the effect of the Coulomb potential on the rescattering electronic motion is key to the understanding of the apparent peak shift in the wavelength-dependence of the HHG yield.

\section{Conclusions}
\label{sec:conclusions}
Using full numerical solutions of the time-dependent Schr\"odinger equation, we have found that the fundamental wavelength dependence of HHG with few-cycle pulses in the single-atom response features surprisingly strong oscillations on fine wavelength scales with modulation periods as small as 6 nm in the mid-infrared regime near $\lambda = 2\, \mu$m. 
Thus, even a slight change in fundamental wavelength leads to strong variations in the HHG yield.
This fine-scale rapid variation is the consequence of the interference of several rescattering trajectories with long excursion times, confirming the significance of multiple returns of the electron wavepacket \cite{ar:tate_scaling07}. 

The present oscillations are closely related to similar regular peak-like enhancements of harmonic yield as a function of intensity $I_0$ \cite{Toma1999JPB, Zair2008PRL,Borca2002PRA,Milosevic2002PRA}, previously discussed in connection with channel closings. Our analysis of the simultaneous wavelength-intensity-dependence has revealed that the spacing between adjacent peaks (expressed in terms of the channel closing number $R$) is very accurately given by $\delta R=1$ over a wide range of $\lambda$ and $I_0$. 
This corresponds to the spacing of adjacent channel closings as predicted by the strong-field approximation (e.g.\  \cite{Milosevic2002PRA}). 
The condition $\delta R=1$ holds only if $R$ is defined with the true ionization potential. 
However, the peak positions are significantly shifted relative to integer values. 
The parametric dependence of the peak shift on the wavelength $\lambda$ and the intensity $I_0$ has been investigated. 
Our analysis shows that this peak shift can be accounted for by the effects of the Coulomb tail on the motion of the returning electron. 

\begin{acknowledgments}
The work was supported by the Austrian ``Fonds zur F\"orderung der wissenschaftlichen Forschung'', under grant no. FWF-SFB016 ``ADLIS''.
K.S.\ also aknowledges support by the IMPRS-APS program of the MPQ (Germany). 
K.L.I.\ gratefully acknowledges financial support by the Precursory Research for Embryonic Science and Technology (PRESTO) program of the Japan Science and Technology Agency (JST) and by the Ministry of Education, Culture, Sports, Science, and Technology of Japan, Grant No. 19686006. 
K.L.I.\ would like to thank P.\ Sali\`eres, T.\ Auguste, and H. Suzuura for illuminating discussions. 
\end{acknowledgments}

\end{document}